\documentclass[aps,prd,preprintnumbers,groupedaddress,nofootinbib,twocolumn]{revtex4}
\pdfoutput=1

\usepackage{graphicx}
\usepackage{latexsym}
\usepackage{amsfonts}
\usepackage{amssymb}
\usepackage{amsmath}
\usepackage{slashed}
\usepackage{array}
\usepackage{feynmp}
\usepackage{url}
\usepackage{caption}
\begin{document}
\title{A Spin-Dependent Interpretation for Possible Signals of Light Dark Matter}

\author{Matthew R.~Buckley$^{1}$ and W.~Hugh Lippincott$^1$}
\affiliation{$^1$Center for Particle Astrophysics, Fermi National Accelerator Laboratory, Batavia, IL 60510, USA}
\preprint{FERMILAB-PUB-13-190-A-AE}
\date{\today}

\begin{abstract}
Signals broadly compatible with light ($7-10$~GeV) dark matter have been reported in three direct detection experiments: CoGeNT, DAMA/LIBRA, and CDMS-II silicon. These possible signals have been interpreted in the context of spin-independent interactions between the target nuclei and dark matter, although there is tension with null results, particularly from xenon-based experiments. In this paper, we demonstrate that the CoGeNT and CDMS-II silicon results are also compatible assuming a spin-dependent neutron interaction, though this is in tension with xenon-based experiments and PICASSO. The tension with the null results from XENON100 and XENON10 is approximately the same as for the spin-independent coupling. All three experimental signals can be made compatible through a combination of spin-dependent interactions with both the proton and neutron, although such a scenario increases the conflict with the null results of other experiments. 
\end{abstract}


\maketitle

With three events in their silicon detectors, CDMS-II \cite{Agnese:2013rvf} joins DAMA/LIBRA \cite{Bernabei:2010mq} and CoGeNT \cite{Aalseth:2011wp,Aalseth:2010vx} in reporting signals that appear to be compatible with light dark matter with a mass in the $7-10$~GeV range. These possible signals of dark matter are interpreted in the context of spin-independent (SI) interactions with nucleons. The spin-dependent (SD) interaction has not been widely considered as a viable alternative for the combination of experimental results, though it has been discussed in the context of the CoGeNT \cite{Kopp:2009qt} and DAMA/LIBRA data \cite{Savage:2004fn}. 

For low momentum transfer, both SD and SI interactions are coherent across the nucleus. This results in a cross section proportional to nucleon number squared for SI interactions. However, for most isotopes the spin of the nucleons are paired, and so the SD interaction does not get an equivalent boost from large nuclei. As a result, most experiments are less sensitive to such interactions. By mass, most isotopes of germanium and silicon have zero spin: germanium has $\sim 8\%$ of nuclei sensitive to SD interactions, while silicon has only $5\%$. CDMS-II contained both germanium (CDMS-Ge) and silicon (CDMS-Si) targets, while CoGeNT is purely germanium. DAMA/LIBRA is a crystal of sodium iodide, both of which consist of nuclei with net spin.  Xenon has many common isotopes with non-zero spin, and the experiments based on this element have reported only limits on dark matter interactions. Thus, one might be led to the conclusion that a cross-section large enough to account for the CDMS-Si result would be incompatible with that required for CoGeNT and DAMA/LIBRA and firmly ruled out by the XENON10 \cite{Angle:2011th} and XENON100 SD limits \cite{Aprile:2013doa}. Though the null results from other experiments are in conflict with these signals, the level of tension appears roughly equivalent to that between the CoGeNT/CDMS-Si SI region of interest and the XENON100 bound.

The spectrum of dark matter events that would be seen in a direct detection experiment is a function of the (theoretically assumed) dark matter microphysics, measured nuclear properties of the target element, and the distribution of dark matter in the local Galaxy (which must be derived from astrophysics, with considerable uncertainties). The differential rate with respect to recoil energy is
\begin{equation}
\frac{dR}{dE_R}  = N_T \frac{\rho_\chi}{m_\chi} \int_{|\vec{v}|>v_{\rm min}} d^3v \frac{f(\vec{v})}{v} \frac{d\sigma}{dE_R}
\end{equation}
This factors out the local dark matter density $\rho_\chi$, dark matter mass $m_\chi$, dark matter velocity distribution $f(\vec{v})$, and the number of target nuclei $N_T$ from the differential dark matter-nucleon cross section $d\sigma/dE_R$, which depends on the particle physics of the dark matter-nucleus interaction. Assuming a particular velocity distribution, the various dark matter direct detection experiments can directly compare the values of integrated $\sigma$ to which they are sensitive. The experiments typically assume a Maxwell-Boltzmann velocity distribution, though it should be noted that the true velocity distribution is expected to deviate from this ansatz, especially at high velocities \cite{Kuhlen:2012fz}.\footnote{See Refs.~\cite{Fox:2010bz,Fox:2010bu,Gondolo:2012rs,HerreroGarcia:2012fu,HerreroGarcia:2011aa} for an alternative parametrization that removes the astrophysical uncertainties.} For a given cross section $\sigma$, the measured rate is also independent of whether the cross section arises from SI or SD interactions. This allows us to convert published SI cross sections into equivalent SD ones.

\begin{figure}[ht]
\includegraphics[width=0.9\columnwidth]{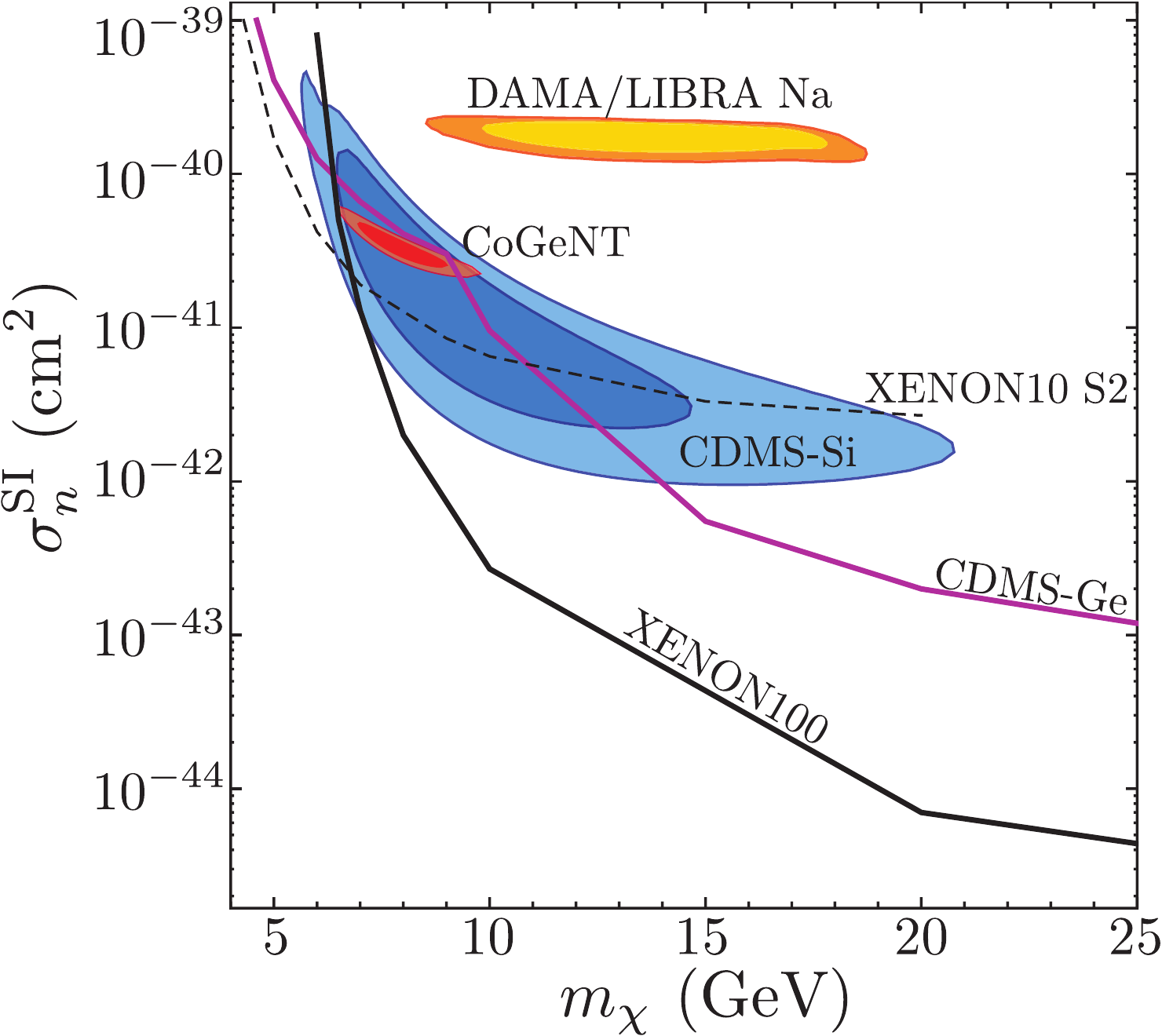}
\caption{Spin-independent nucleon cross section as a function of dark matter mass $m_\chi$, assuming isospin-conserving interactions. Upper bounds are reported by XENON100 \cite{Aprile:2012nq} (black line), XENON10 S2 \cite{Angle:2011th} (black dashed line) and CDMS-II germanium \cite{Ahmed:2009zw} (purple line). Regions compatible with the events seen in CoGeNT \cite{Aalseth:2011wp,Aalseth:2010vx,Kelso:2011gd}, CDMS-Si \cite{Agnese:2013rvf}, and DAMA-LIBRA (assuming $Q_{\rm Na} = 0.25$~\cite{Kelso:2011gd,Collar:2013gu}) are shown in red (90\% and 99\% contours), blue (68\% and 90\% contours),  and yellow (90\% and 99\% contours) respectively.  \label{fig:SI}}
\end{figure}

For a target nucleus with atomic number $Z$ and mass number $A$, the elastic SI cross section at recoil energy $E_R$ can be written in terms of a dark matter-proton cross section $\sigma_p^{\rm SI}$, the dark matter-nucleus reduced mass $\mu$, the dark matter-proton reduced mass $\mu_{p}$, and the proton and neutron couplings $f_p$ and $f_n$:
\begin{equation}
\sigma_{\rm SI}=  \frac{\mu^2}{\mu_p^2}\frac{[f_p Z + f_n(A-Z)]^2}{f_p^2} \sigma^{\rm SI}_p. \label{eq:SIsigma}
\end{equation}
A nuclear form factor $F(q^2)$ has also been factored out of the cross section to isolate the unknown particle physics component of the dark matter interaction, which comes into play in both the proton cross section $\sigma_p^{\rm SI}$ and the couplings $f_p$ and $f_n$. The experimental bounds on the elastic cross section typically assume $f_p = f_n$ (isospin conserving). Relaxing this assumption can decrease the sensitivity to dark matter for specific experiments. For example, if $f_n = -0.7f_p$, xenon-based experiments will have extremely low sensitivity to dark matter compared to the germanium- and silicon-based targets  \cite{Kurylov:2003ra,Giuliani:2005my,Frandsen:2013cna,DelNobile:2013cta}. Assuming an isospin-conserving SI interaction, the CDMS-Si result is compatible with a dark matter-nucleon cross section of approximately $\sim 2\times 10^{-41}$~cm$^2$. In Fig.~\ref{fig:SI}, we plot the SI bounds from XENON-100 \cite{Aprile:2012nq}, XENON10 S2-only analysis~\cite{Angle:2011th} and CDMS-Ge \cite{Ahmed:2009zw}, as well as the regions compatible with the reported events in CoGeNT \cite{Aalseth:2011wp,Aalseth:2010vx,Kelso:2011gd}, CDMS-Si \cite{Agnese:2013rvf} , and DAMA-LIBRA \cite{Bernabei:2010mq} (assuming a quenching factor of $Q_{\rm Na} = 0.25$, as in Ref.~\cite{Collar:2013gu}). CRESST-II also reports excess events in broad agreement with light dark matter \cite{Angloher:2011uu,Kopp:2009qt}, though a possible unresolved background could impact these results \cite{Kuzniak:2012zm}. As the light nuclei in CRESST-II (oxygen and calcium) do not have significant abundances of non-zero spin isotopes, we do not include this result in our study.

A spin dependent interaction couples to the total spin of the nucleus. However, as nucleon spins are typically paired, there is no large boost to the nuclear SD cross section comparable to the $A^2$ enhancement that SI interactions receive. Instead, the cross section couples to the total nuclear spin $J$, which is zero unless there is an unpaired nucleon. Even for such cases, $J$ is usually small. The elastic SD nuclear cross section is
\begin{equation}
\sigma_{\rm SD}(E_R) = \mu^2 [a_p\langle S_p\rangle + a_n\langle S_n\rangle]^2 \frac{J+1}{J}. \label{eq:SDsigma}
\end{equation}
Here, $a_p$ and $a_n$ encode the proton and neutron couplings, respectively, and depend on the assumed dark matter microphysics. $\langle S_P\rangle$ and $\langle S_n\rangle$ are the spin expectation values for the proton and neutron groups in the nucleus.\footnote{An alternative formulation, in terms of isoscalar and isovector couplings, is equivalent. See Refs.~\cite{Kopp:2009qt,Bednyakov:2006ux}.} To compare limits and signals across experiments, $\sigma_{\rm SD}$ can be converted into an effective proton or neutron cross section using
\begin{equation}
\sigma^{\rm SD}_{p,n} = \frac{3}{4} \frac{J}{J+1} \frac{\mu_{p,n}^2}{\mu^2} \frac{\sigma_{\rm SD}}{\langle S_{p,n}\rangle^2}. \label{eq:SDsigmaN}
\end{equation}
This assumes that the interactions proceeds solely through $a_p$ or $a_n$, but not both.

In Table~\ref{tab:elements}, we list the the isotopes used in the relevant dark matter direct detection experiments that are sensitive to spin dependent interactions, along with their abundance, nuclear spin, and $\langle S_{p,n} \rangle$ values. Only a small fraction of the silicon and germanium targets are relevant for SD scattering. It should be emphasized that the expectation values of the nuclear spins are extracted from theoretical calculations. It is therefore not implausible that the true values of $\langle S_{p,n}\rangle$ differ from the ones used in this work (see, for example, the change in xenon $\langle S\rangle$ calculated using different models in Refs.~\cite{Ressell:1997kx,Menendez:2012tm}).  

\begin{table}[th]
\begin{tabular}{|c||c|c|c|c|}
\hline
Isotope & Abundance & $~J~$ & $\langle S_p\rangle$ & $\langle S_n\rangle$ \\ \hline \hline
$^{19}$F \cite{Tovey:2000mm} & 100\% & $\tfrac{1}{2}$ & 0.441 & -0.109  \\ \hline
 $^{23}$Na \cite{Ressell:1997kx} & 100\% & $\tfrac{3}{2}$ & 0.248 & 0.020 \\ \hline
 $^{29}$Si \cite{Tovey:2000mm} & 4.7\% & $\tfrac{1}{2}$ & -0.002 & 0.130 \\ \hline
$^{73}$Ge \cite{Tovey:2000mm} & 7.8\% & $\tfrac{9}{2}$ & 0.030 & 0.378 \\ \hline
$^{127}$I \cite{Ressell:1997kx}& 100\% & $\tfrac{5}{2}$ & 0.309 & 0.075 \\ \hline
$^{129}$Xe \cite{Menendez:2012tm}& 26.4\% & $\tfrac{1}{2}$ & 0.010 & 0.329 \\ \hline
$^{131}$Xe \cite{Menendez:2012tm} & 21.2\% & $\tfrac{3}{2}$ & -0.009  & -0.272 \\ \hline
\end{tabular}
\caption{The isotopes in direct detection experiments sensitive to SD interactions, along with their relative abundance, spin $J$, and theoretical $\langle S_p\rangle$ and $\langle S_n\rangle$ values \cite{Giuliani:2005bd}. The references for the spin expectation values are included in the first column. \label{tab:elements}}
\end{table}

\begin{figure}[t]
\includegraphics[width=0.9\columnwidth]{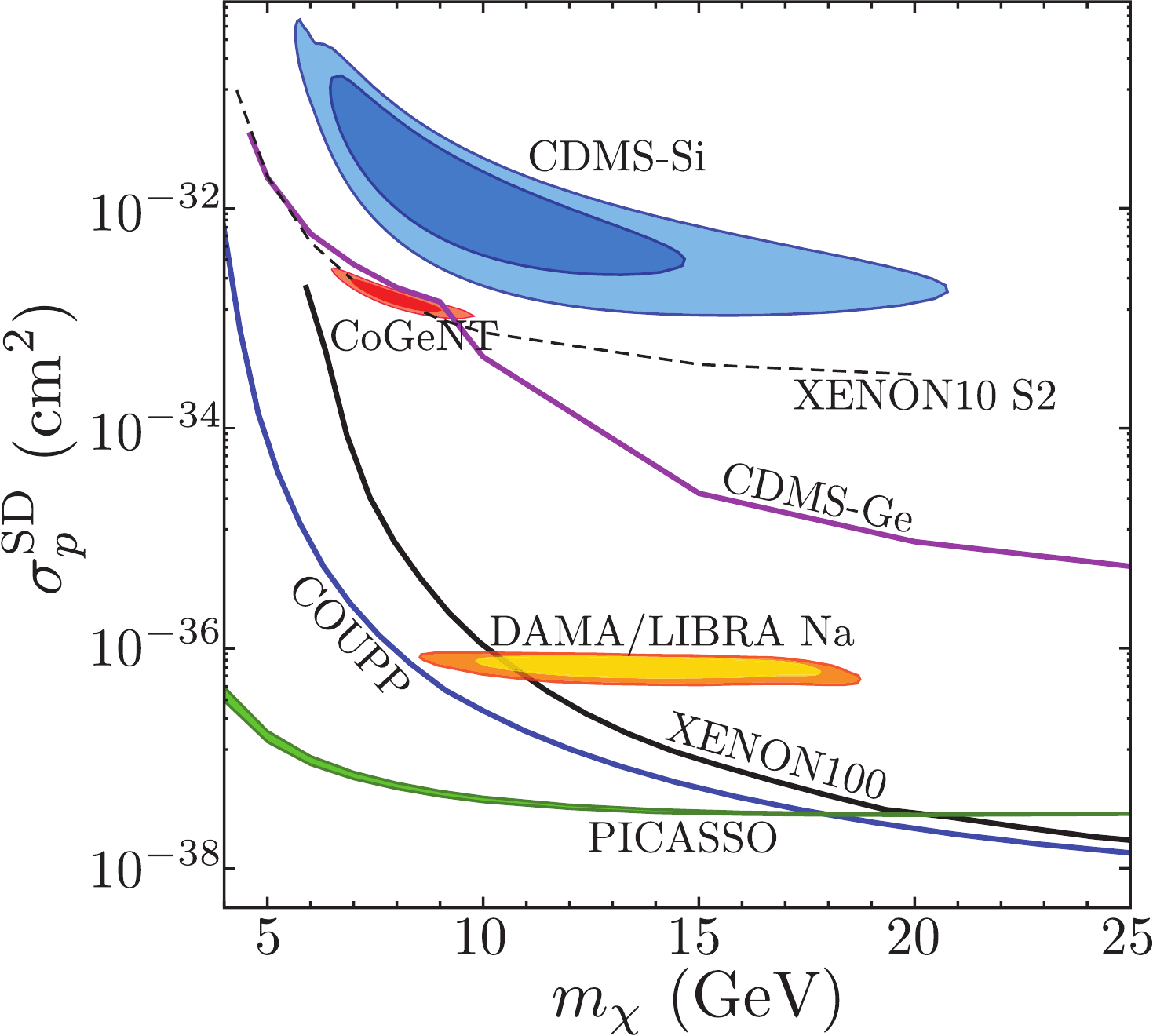}
\caption{Spin-dependent proton cross section as a function of dark matter mass $m_\chi$, assuming interactions solely with protons. The DAMA/LIBRA region assumes 100\% scattering with sodium.  PICASSO and COUPP limits are also shown, with all other labeling as in Fig.~\ref{fig:SI}.  \label{fig:SDp}}
\end{figure}

\begin{figure}[t]
\includegraphics[width=0.9\columnwidth]{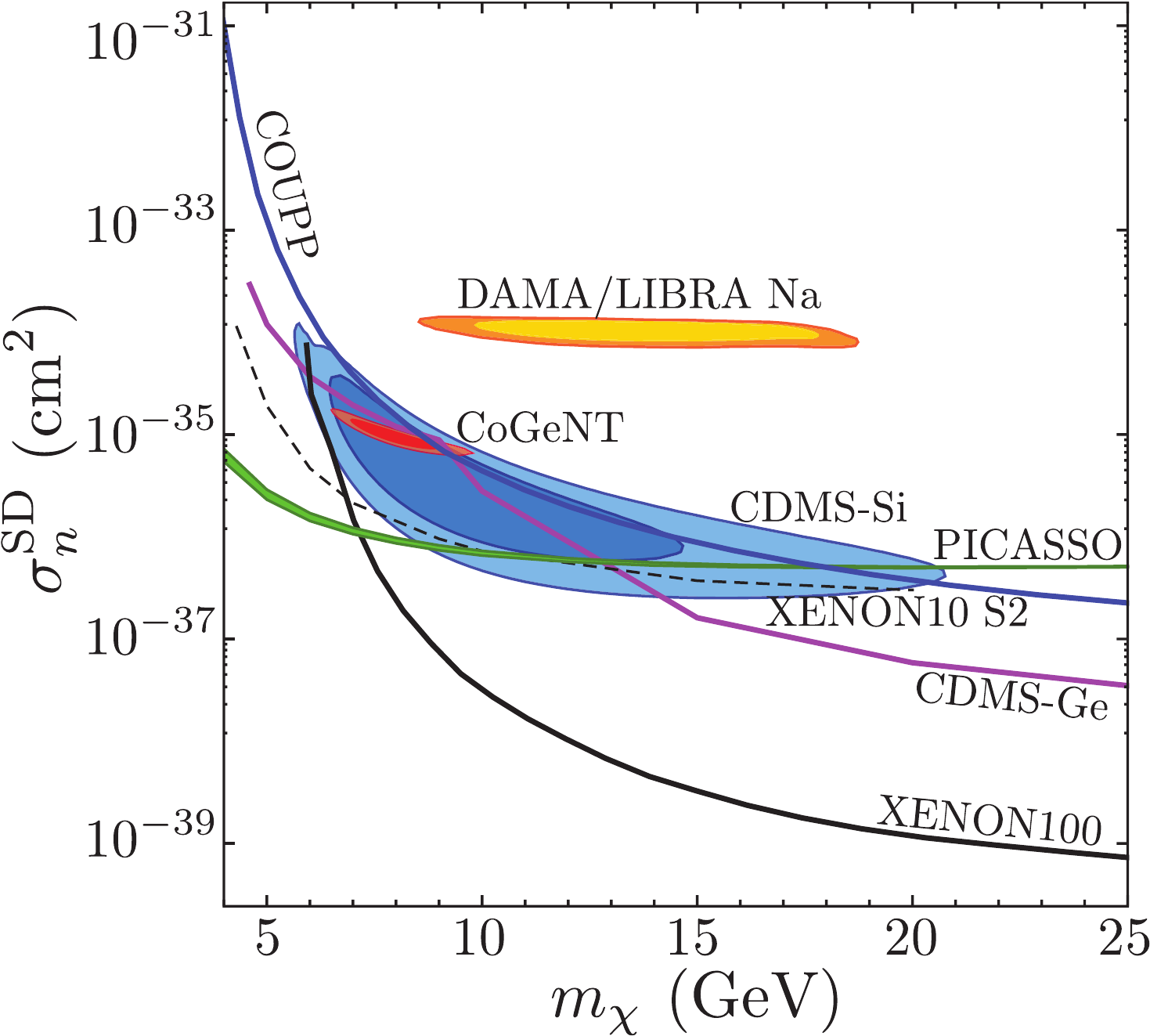}
\caption{Spin-dependent neutron cross section as a function of dark matter mass $m_\chi$, assuming interactions solely with neutrons. Labeling as in Fig.~\ref{fig:SDp}.   \label{fig:SDn}}
\end{figure}

Working with the available values of $\langle S_{p,n}\rangle$, we use Eqs.~\eqref{eq:SIsigma}-\eqref{eq:SDsigmaN} to convert the published spin independent nucleon cross section limits and signal regions from CoGeNT, CDMS-Si, CDMS-Ge, and DAMA/LIBRA into an equivalent spin dependent cross section assuming coupling to either protons or neutrons. For DAMA/LIBRA, there are two regions in the mass vs.~SI cross section plane that are consistent with the observed modulation. The low mass region visible in Fig.~\ref{fig:SI} is the result of scattering from the lighter sodium atoms, and we use the sodium nuclear properties to translate into a SD region. The results are shown in Figs.~\ref{fig:SDp} and \ref{fig:SDn}, along with the published XENON100 bounds from Ref.~\cite{Aprile:2013doa}.\footnote{As a cross-check of our conversion from SI to SD cross sections, we verified that we reproduce the SD results of Refs.~\cite{Aprile:2013doa} and \cite{Akerib:2005za} using the published SI XENON100 \cite{Aprile:2012nq} and SI CDMS-Ge data \cite{Ahmed:2009zw}. Although we find a weaker bound on $\sigma_n^{\rm SI}$ than reported by XENON100, we plot their published results in Fig.~\ref{fig:SDn}.} Fig.~\ref{fig:SDp} also shows the limits on SD-proton coupling from the PICASSO \cite{Archambault:2012} and COUPP collaborations~\cite{Behnke:2012}. For these two experiments, we convert the SD proton cross sections to an equivalent $\sigma_n^{\rm SD}$ using the values of $\langle S_{p,n}\rangle$ for fluorine in Table~\ref{tab:elements}. The results are shown in Fig.~\ref{fig:SDn}.


As can be seen from these figures, the best fit regions of CDMS-Si and CoGeNT coincide when the scattering proceeds exclusively through neutrons, and not if it goes through protons only. Both the XENON100 and PICASSO bounds are in conflict with the signal regions in the neutron scattering. As with SI scattering, one could appeal to possible deviations from the assumed Maxwell-Boltzmann distribution of dark matter \cite{Mao:2013nda} or accidental cancellations in xenon through isospin violating couplings \cite{Kurylov:2003ra,Giuliani:2005my,Frandsen:2013cna,DelNobile:2013cta} in order to relax this tension. Note that to cancel the XENON100 bounds through isospin violation, we would require $a_p/a_n \sim -30$, and PICASSO would require $a_p/a_n \sim 4$. It is unclear whether either of these scenarios can be realized in realistic models of dark matter while avoiding all all other experimental bounds. The possible issues concerning XENON100 sensitivity to low energy recoils would also be relevant to the SD case as they are in SI (see {\it e.g.}~Refs.~\cite{Collar:2011wq,Horn:2011wz,Manalaysay:2010mb,Manzur:2009hp,Plante:2011hw,Hooper:2013cwa}). Beyond these uncertainties which are present in both SD and SI interpretations of the experimental results, it is possible that additional uncertainties from the calculation of $\langle S_{p,n}\rangle$ are relevant. Further work on the astrophysical and experimental uncertainties is necessary to determine whether all results can be brought into agreement.

The DAMA/LIBRA regions appear to be inconsistent with the CoGeNT and CDMS-Si regions in both the neutron- and proton-only scattering. However, the DAMA/LIBRA regions are low compared to CoGeNT/CDMS-Si for $\sigma^{\rm SD}_p$ and high when the scattering is through $\sigma^{\rm SD}_n$. A dark matter coupling to both proton and neutrons can move these regions into closer alignment. For example, $a_n = a_p$ brings all the best-fit regions of all three experiments into close agreement. However such a model would be in conflict with the strong bounds on SD proton scattering from XENON100, PICASSO, and COUPP.

Spin dependent interactions require much larger cross sections with nucleons than required in SI scattering. If we assume that this interaction is mediated by an effective operator \cite{Goodman:2010ku}, the collider-based searches for monojets \cite{Beltran:2010ww,Fox:2011pm,Goodman:2010ku,ATLAS:2012ky,Chatrchyan:2012me} and mono-$W/Z/\gamma$ \cite{Aad:2012fw,Chatrchyan:2012tea,Bai:2012xg,Aad:2012awa,Carpenter:2012rg,Zhou:2013fla} place significant bounds on the mass scale suppressing such interactions. It is possible that such constraints may require any model that attempts to explain the possible light dark matter signal in terms of SD scattering to include either dark sectors or light mediators. Further bounds on dark matter with large cross sections also exist from dark matter capture and annihilation in the Sun, which can constrain the final states that such dark matter could annihilate into \cite{Desai:2004pq,Fitzpatrick:2010em}. More study is required to determine which, if any, effective operator models are compatible with the SD interpretation.

The nuclear properties that control a target's sensitivity to SD scattering differ greatly from element to element, and one would not expect that signals compatible with similar SI cross sections in germanium and silicon would also give good agreement in SD. However, we find a dark matter candidate with mass $\sim 7-10$~GeV and neutron scattering of $\sigma^{\rm SD}_n\sim 10^{-35}$~cm$^2$ gives a good fit to both CoGeNT and CDMS-Si, though this region is in conflict with the PICASSO, XENON100, and XENON10 null results. The level of tension appears equivalent to the tension between XENON100 and CoGeNT/CDMS-Si seen in the SI case. Adding a proton scattering can bring the DAMA/LIBRA signal into agreement as well, though this seems disfavored by a number of other SD experiments.

Whether the anomalous events seen in DAMA/LIBRA, CoGeNT, and CDMS-II are due to dark matter or some unknown background is still unclear, and further investigation from multiple experiments is required. However, as we attempt to interpret such signals, theorists and experimentalists must remain open to all possible scenarios. As we have demonstrated, spin dependent interactions with neutrons give a region of parameter space that appears to explain both CoGeNT and CDMS-Si. While tension exists with current experiments, the possibility of a SD origin of these results should be kept in mind as we await new results from LUX \cite{Akerib:2012ys}, CoGeNT, SuperCDMS \cite{Schnee:2005pj}, and COUPP with a C$_3$F$_8$ target.


\bibliographystyle{apsrev}
\bibliography{spindependent5}
\end{document}